\documentclass[manuscript]{acmart}

\AtBeginDocument{%
  \providecommand\BibTeX{{%
    \normalfont B\kern-0.5em{\scshape i\kern-0.25em b}\kern-0.8em\TeX}}}

\setcopyright{acmcopyright}

\copyrightyear{2024} 
\acmYear{2024} 
\setcopyright{acmlicensed}\acmConference[CHI '24]{Proceedings of the CHI Conference on Human Factors in Computing Systems}{May 11--16, 2024}{Honolulu, HI, USA}
\acmBooktitle{Proceedings of the CHI Conference on Human Factors in Computing Systems (CHI '24), May 11--16, 2024, Honolulu, HI, USA}
\acmDOI{10.1145/3613904.3642261}
\acmISBN{979-8-4007-0330-0/24/05}

\acmSubmissionID{7662}

\usepackage{graphicx}
\usepackage{CJKutf8}

\def\markup{0}
\if\markup1
\usepackage[normalem]{ulem}
  \definecolor{myblue}{rgb}{0,0,0.75}
  \newcommand{\rv}[1]{{\leavevmode\color{myblue}#1}}
  \newcommand{\st}[1]{{\sout{#1}}}
\else
  \newcommand{\rv}[1]{#1}
\newcommand{\st}[1]{}
\newcommand{\sout}[1]{}
\fi

\begin{document}

\title[Design Makeup Residue Visualization System for Chinese Traditional Opera (Xiqu) Performers]{``It Is Hard to Remove from My Eye'': Design Makeup Residue Visualization System for Chinese Traditional Opera (Xiqu) Performers
}


\author{Zeyu Xiong}
\affiliation{
  \institution{Computational Media and Arts Thrust}
  \institution{The Hong Kong University of Science and Technology (Guangzhou)}
  \city{Guangzhou}
  \country{China}
}
\email{zxiong666@connect.hkust-gz.edu.cn} 

\author{Shihan Fu}
\affiliation{
  \institution{Computational Media and Arts Thrust}
  \institution{The Hong Kong University of Science and Technology (Guangzhou)}
  \city{Guangzhou}
  \country{China}
}
\email{sfu663@connect.hkust-gz.edu.cn} 

\author{Yanying Zhu}
\affiliation{
  \institution{Data Science and Analysis Thrust}
  \institution{The Hong Kong University of Science and Technology (Guangzhou)}
  \city{Guangzhou}
  \country{China}
}
\email{yzhu367@connect.hkust-gz.edu.cn} 

\author{Chenqing Zhu}
\affiliation{
  \institution{Internet of Things Thrust}
  \institution{The Hong Kong University of Science and Technology (Guangzhou)}
  \city{Guangzhou}
  \country{China}
}
\email{czhu032@connect.hkust-gz.edu.cn} 

\author{Xiaojuan Ma}
\affiliation{
  \institution{Department of Computer Science and Engineering}
  \institution{The Hong Kong University of Science and Technology}
  \country{Hong Kong SAR, China}
}
\email{mxj@cse.ust.hk} 

\author{Mingming Fan}
\orcid{0000-0002-0356-4712}
\affiliation{%
  \institution{Computational Media and Arts Thrust}
  \institution{The Hong Kong University of Science and Technology (Guangzhou)}
  \city{Guangzhou}
  \country{China}
}
\affiliation{%
  \institution{Division of Integrative Systems and Design \& Department of Computer Science and Engineering}
  \institution{The Hong Kong University of Science and Technology}
  \country{Hong Kong SAR, China}
}
\authornote{Corresponding Author}
\email{mingmingfan@ust.hk}

\renewcommand{\shortauthors}{Xiong, et al.}

\begin{abstract}
   Chinese traditional opera (Xiqu) performers often experience skin problems due to the long-term use of heavy-metal-laden face paints. To explore the current skincare challenges encountered by Xiqu performers, we conducted an online survey (N=136) and semi-structured interviews (N=15) as a formative study. We found that incomplete makeup removal is the leading cause of human-induced skin problems, especially the difficulty in removing eye makeup. Therefore, we proposed \textit{EyeVis}, a prototype that can visualize the residual eye makeup and record the time make-up was worn by Xiqu performers. We conducted a 7-day deployment study (N=12) to evaluate \textit{EyeVis}. Results indicate that \textit{EyeVis} helps to increase Xiqu performers' awareness about removing makeup, as well as boosting their confidence and security in skincare. Overall, this work also provides implications for studying the work of people who wear makeup on a daily basis, and helps to promote and preserve the intangible cultural heritage of practitioners.
\end{abstract}



\begin{teaserfigure}
  \includegraphics[width=\textwidth]{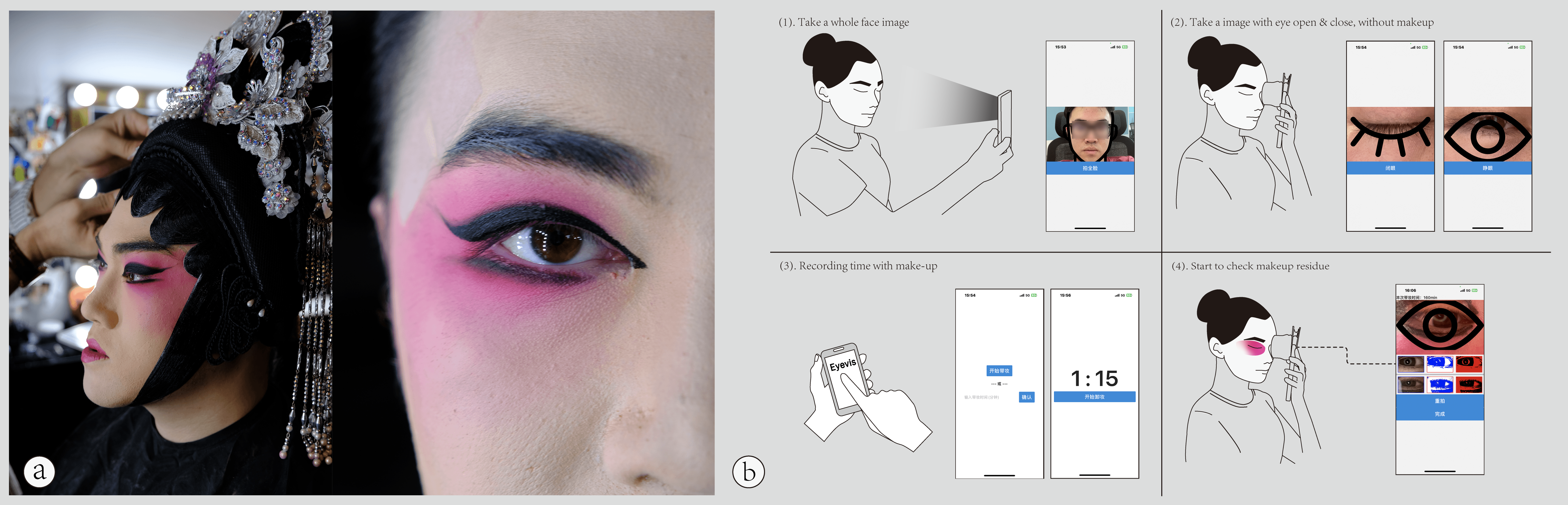}
  \caption{Overview of Makeup Residue Visualization System (The photo is used with consent). (a) Chinese Traditional Opera Makeup, (1) - (4): Sample Usage of the system, step (1): Take an image of the whole face, step (2): Take images of eyes (open \& close), step (3): Recording Makeup Wearing Time, and step (4): Eye Makeup Visualization.}
  \Description{Figure 1 shows the overview of the makeup residue visualization system. Left shows a sample picture of Chinese Traditional Opera Makeup, Right shows the sample usage of the system, which includes four main steps:  (1): Take an image of the whole face, (2): Take images of eyes (open \& close), (3): Recording Makeup Wearing Time, and (4): Eye Makeup Visualization.}
  \label{fig:teaser}
\end{teaserfigure}


\maketitle

\section{Introduction}
\label{sec: intro}

Stage performers have a high risk of suffering from occupational health problems~\cite{hinkamp2017occupational}. Especially for the performing arts which require high-quality visual effects, actors might need to wear heavy makeup frequently to achieve such effects and are therefore heavily exposed to cosmetics~\cite{farm1995}. Wearing makeup for a long period of time as well as inappropriate and incomplete makeup removal can easily cause skin irritation and a host of other problems, especially if the makeup contains heavy metals~\cite{wang2020heavy}. 
One of the representative user groups is Chinese traditional opera (Xiqu) performers. Xiqu is one type of the most significant intangible cultural heritage (ICH) under the United Nations Educational, Scientific and Cultural Organization (UNESCO) list \cite{Browseth50:online}. As intangible cultural practitioners, Xiqu performers have made great contributions to society. According to the investigation from the Ministry of Culture (PRC) \cite{92:online} by 2018, there are 348 genres of Xiqu and over 10,278 troupes registered, and there are over 300,000 Xiqu performers performing actively on stage, based on the survey of China Theater Association \cite{2:online} by 2019.

Xiqu is also a form of musical stage art that highly demands visual effects~\cite{liu1997art, corson2019stage, wang2013research}. Specifically, thick face paints are usually used with various colors in terms of different roles of play (e.g. Sheng, Dan, Jing, Mo, Chou~\cite{riley1997chinese}) to achieve considerable visual effects \cite{eldridge2015face} and expression of emotions~\cite{zhou20073d}. However, face paint consists of various types of heavy metals \cite{attard2022heavy, contado2012new, borowska2015metals}, and the long wear of face paint is extremely harmful to the skin \cite{wang2020heavy}. It is usually used by Xiqu performers for about a quarter of the year on average~\cite{wang2023insights}. 
Aside from the necessity of the performance, wearing make-up for a long time can also be due to incomplete makeup removal, which might lead to skin problems such as pore clogging, increased oil secretion, skin allergies, and acne vulgaris \cite{kohl2002allergic}. Therefore, the skin health of Xiqu performers caused by face paint makeup needs more attention. To reduce the hassle of applying and removing traditional makeup, researchers have proposed new materials for makeup~\cite{kao2016chromoskin, 10.1145/2663806.2663823} and interactive makeup methods~\cite{10.1145/3277452}. However, little is known about preventing skin damage during the makeup removal process from the user's perspective. To this end, we propose a research question (RQ1): \textbf{What are the practices and challenges of making-up removal among Xiqu performers?}

To answer RQ1, we conducted a formative study in which we first performed a survey that understands the fundamental data of Xiqu performers' makeup removal process. Among 136 responses, we selected 15 representative participants to conduct a semi-structured interview to understand the detailed challenges and practices. We open-coded the results and found that incomplete makeup removal is the main reason, and the eye makeup is the hardest part to remove, and we derived six design considerations (DC). Based on the DCs, we further designed and implemented \textit{EyeVis}, 
an interactive system that is tailored for Xiqu performers to visualize the eye makeup residue during the removal process, and track the trend of makeup-wearing time, consisting of a mobile app with a fill light \& a camera lens magnifier \& eye shields. To our knowledge, \textit{EyeVis} is the first ever tool designed for eye makeup residue visualization. We then used it to investigate the effectiveness and user experience of these features by answering the second research question
(RQ2): \textbf{What features and functionalities would Xiqu performers find valuable in a system designed to aid in eye makeup removal?} To answer RQ2, we conducted a 7-day deployment study where 12 Xiqu performers participated simultaneously. During the deployment study, participants were asked to complete makeup removal tasks every day and participated in the final day interview. We open-coded the data and derived our findings by data analysis. The results showed the usability of~\textit{EyeVis}, including the visualization methods to improve the efficiency of the makeup removal process from multiple perspectives and increase the awareness of the Xiqu performers. Finally, we discussed potential generalization directions for our work in terms of wider visualization area of skin,  wider audiences, and wider usage for skin data collection.

In summary, we made the following contributions:

\begin{itemize}
    \item We investigated common practices and challenges for the making-up removal process of Xiqu performers through a formative study including an online survey and semi-structured interviews, and we proposed six informative design considerations for the future design of similar skincare systems. 
    \item Informed by the formative study, we designed and implemented an interactive prototype that allows Xiqu performers to see the makeup residue clearly through the visualization system, and see the makeup wearing time by a weekly trend.
    \item We conducted a 7-day deployment study with a community of Xiqu performers to understand how they used and perceived these features and presented design implications. To our knowledge, this is the first prototype that has been designed and evaluated for makeup residue visualization.
\end{itemize}

\section{Related Work}

\subsection{Xiqu Performers Suffer from Skin Problems}

Standard face paint products for professionals are subjected to frequent random market controls in Western countries, in accordance with European Union (EU) cosmetic product regulations~\cite{keck2015}. However, face paint, which is often used by Chinese traditional opera (Xiqu) performers, tends to be unregulated in China~\cite{michalek2019systematic}. As such, their chemical composition and the degree of protection they provide to the face are questionable. As a result of wearing face paint for years and years, the skin condition of Xiqu performers is generally poor. Wang et al.~\cite{facedermatitis1999} found that adverse reactions to face paint were present in 67.7 \% of the performers, with more heavy reactions in the Dan roles (36.7 \% of those with reactions) by investigating 127 Xiqu performers from Beijing Opera Theater. There were other manifestations of skin damage, such as seborrheic dermatitis, in 73 \% of the performers. Wang et al.~\cite{wang2023insights} conducted a questionnaire survey in which 259 Xiqu performers participated, and 53\% of them reported skin diseases like skin erythema and dermatitis because of the long-term use of face paint. Besides, The condition of Xiqu's skin may get worse over the years of the performer's career. For example, Cao et al.~\cite{2018jingju} reported 2 cases of skin melanoma among Xiqu performers at the Beijing Opera Theater in 2016, and these performers both have over 30 years of experience in the field. Based on the reports and surveys mentioned above, we can find that stage performers, especially  Xiqu performers do have potentially serious skin problems. Despite many studies have focused on how to lessen the toxicity of cosmetics from a chemical and medical perspective~\cite{wang2016cultural, wang2023insights, WANG2023164374}, there is currently little attention paid to the skincare of stage performers, especially Xiqu performers, in the HCI community.

To reduce the hassle of applying and removing traditional makeup, some researchers are working on new materials and interactive makeup. For example, Kao et al.~\cite{kao2016chromoskin} introduced a prototype of an interactive eyeshadow tattoo consisting of thermochromic pigments that can be activated by electronic or ambient temperature conditions. Treepong et al.~\cite{10.1145/3277452} developed an augmented reality system that scans the human face and projects the makeup effects on the face to enhance makeup creativity. \citet{10.1145/2663806.2663823} proposed iMake, an eye makeup design and printing system that utilizes ink to attach to eyes. Another alternative to traditional makeup is using masks, which are widely used in Western opera. However, Tseng and Lin~\cite{tseng2013comparison} have investigated the difference between Xiqu makeup and masks used in Western opera, although painted faces and masks share some common functions, face paint represents the essence of Xiqu and is dissonant to be replaced by masks. Therefore, due to the uniqueness of Xiqu makeup and its irreplaceability, none of these existing works could be used in a Xiqu performer's practice, and we have yet to find a way to achieve the same visual effect without the use of face paint.

\subsection{From Human's Perspective: Incomplete Makeup Removal}

With the known risks associated with the chemicals in face paint, it is still essential to wear face paint to achieve considerable visual effect and express inherited emotion on stage~\cite{wang2016cultural}. As most skin diseases are regarded as less severe or life-threatening than other organ problems, the importance of makeup removal is often overlooked. However, as skin problems are frequently visible to the public, they can have a significant impact on a patient's emotional condition~\cite{matsuoka2006}, and even affect Xiqu performers' careers, which may result in them having to stop engaging in opera performance~\cite{farm1995}. Aside from the fact that face paint must be worn for the show, an important human-controllable factor is how thoroughly the makeup is removed \cite{kohl2002allergic}. Due to the ever-evolving chemistry of cosmetics, many of them are waterproof and difficult to remove to prevent blooming~\cite{xing2019category}. If the makeup is not removed completely, it adheres to the surface of the skin and clogs the pores, causing a range of skin problems such as acne, blemishes, inflammation, etc~\cite{park2014allergic}. Especially for Xiqu makeup, heavy makeup is applied to the concave and convex areas of the face (e.g. eye sockets, lips, bridge of the nose), making it more difficult to remove~\cite{lai1972lead}. Although incomplete makeup removal is a crucial factor in influencing skin conditions and can be controlled by Xiqu performers, the extent to which incomplete makeup removal actually causes skin problems for Xiqu performers has not been validated by systematic research.

\subsection{Computer-Aided Skincare Tools in HCI community}

With all the challenges Xiqu performers suffer from wearing face paint, skincare has also become an important part of the post-show experience. Skincare has a strong connection with computer-aided tools~\cite{10181309, asan2021research} in the intersection of medicine and HCI communities, researchers have made contributions to (1) cosmetic product development~\cite{sunkle2021integrated, zhang2021optimization, sharma2021chalpred}, (2) cosmetics usage and makeup tutorial~\cite{10.1145/3491102.3517490, 10.1145/3411764.3445721, 10.1145/3411764.3445364, xiong2023operartistry}, (3) skin physical assessment~\cite{chirikhina2021skin, zegour2023convolutional, 10.1145/3027063.3052754}, (4) skin condition diagnosis~\cite{goldsberry2014visia, wang2021metagenomic, liu2023intelligent, yang2021development}, (5) treatment \& skincare products recommendation~\cite{tian2019smart, huang2018cloud, 10.1145/3123266.3127926, alashkar2017examples}, (6) treatment outcome prediction~\cite{shi2022personalized, bahcceci2021analysis, 10.1145/1753846.1754061}, etc. The above work mainly involves skin imaging technology, which makes contributions to skin diagnosis and treatment by using artificial intelligence to understand various types of image information. However, we found that the focus of these existing works is mostly on the diagnosis and treatment, and how to explore the prevention of skin problems from a human-centered perspective has not yet been researched deeply.

Computer-aided techniques are also being utilized in the HCI community for everyday healthcare in the Chinese context. For example, Ding et al.~\cite{10.1145/3290605.3300435} developed a mobile app that analyzes the face and tongue images for traditional Chinese medicine diagnosis. Wang et al.~\cite{10.1145/3411764.3445432} proposed an AI-powered clinical decision support system in rural areas of China. From our understanding, although the research methodologies among the above works in the Chinese context have reference value, there is no such work in HCI that focuses on skincare for Xiqu performers. Although existing commercial products (e.g. facial cleansers, beauty devices) could help Xiqu performers in skincare, there is still a lack of objective metrics (e.g. augmented makeup residue visualization, skin condition metrics) that can be justified by themselves in their skincare process, various metrics still need industry experts to identify. Therefore, in addition to understanding the problems and challenges encountered by Xiqu performers in the make-up removal process, how to use computer-aided tools to help them directly is also the focus of this work.

\section{Formative Study}

To answer RQ1 about understanding the practices and challenges of making-up removal mentioned in section~\ref{sec: intro}, we conducted a formative study including two parts. First, we collected 136 valid responses via an online survey to investigate the makeup removal process among Xiqu performers. The primary objective of this survey was to gain initial insights into the challenges encountered during the makeup removal process. Second, we conducted semi-structured interviews (N=15) to further explore the Xiqu performers' specific needs. The findings from the formative study showcased the user need for the design of the prototype, which enables the visualization of makeup removal residue during makeup removal and obtains feedback on makeup duration and wearing trends.

\subsection{Online Survey}
To gain deeper insights into the makeup removal process among Xiqu performers, We first designed a comprehensive online survey and distributed it through Xiaohongshu\footnote{\url{https://www.xiaohongshu.com/}}, which is currently one of the most popular social media platforms in China. We posted posters on the platform with specific tags (i.e. \#Xiqu, \#Chinese Opera, \#Makeup, \#Makeup Removal) for the online platform recommendation to targeted users, and we also sent direct messages to accounts that identified as Xiqu performers. \rv{We collected 178 responses and screened 136 valid results by checking the reasonable completion time, the consistency of answers to repetitive questions, and setting some questions regarding professionalism (e.g. choice of make-up tools, order of applying face makeup, mechanism of make-up touch-up after long performances, etc.).} The online survey covered various aspects related to makeup removal, including participants' demographic information (gender, age, years of performing experience, and type of play performed), frequency and duration of wearing face painting, perception of painting's impact on the skin, ease of removing makeup from different facial areas, and skin problems encountered.  


\subsection{Semi-Structured Interviews}
To validate the findings from the online survey and identify other potential user needs, we conducted online interviews with 15 Xiqu performers (5 males and 10 females), with varying years of performing experience (for specific demographic data of participants, see Appendix~\ref{append: demographic}). We recruited participants by contacting participants in the online survey who were interested in participating in our follow-up interviews. Some of the participants who contacted us through the information on our posters also participated in the interviews. \rv{Before the interview, every participant signed a consent form about collecting data for research purposes, and the consent forms passed the university's ethical review.}

The interviews were semi-structured, lasting between 40 and 60 minutes. The interview questions focused on the process participants experienced to remove their makeup, the challenges they encountered, and how they believed face paint affected their skin conditions. The interviews were audio-recorded with the participant's consent, and each participant received a compensation of RMB 100 at the end of the interview.


\subsection{Data Analysis}
Our data consisted of online survey data as well as audio recordings from semi-structured interviews. The audio recordings were transcribed into a text script. To analyze the interview data, we followed the thematic analysis~\cite{oktay2012grounded} method. Four researchers first read through the script to reach an overall understanding, then the open-coding procedure was carried out independently by the two co-authors once they had familiarized themselves with the data. Then, following a weekly discussion, all co-authors in the research team shared individual coding results, discussed interpretations of the data, and achieved consensus on the final coding result. 

\subsection{Findings}

\label{sec: formative-findings}

\subsubsection{Makeup Removal Process}
\paragraph{Challenges in Eye Makeup Removal.}
Many Xiqu performers have expressed that removing eye makeup is the most challenging part of the makeup removal process (N=14). First, in opera performances, emphasis is placed on facial features, and multiple layers of oil-based face paints are applied to the eyes to enhance their prominence. Second, due to the concave structure and presence of wrinkles around the eyes, repeated friction is required to remove the oil-based face paints, which results in a longer time spent on the makeup removal process. (\textit{P5: "The eyeliner in Xiqu requires the oil-based face paints to be applied inside the eyelashes, very close to the eyes, which makes it difficult to clean them completely"}). Besides, the skin in the eye region is more delicate compared to other regions, and there is a risk of getting chemicals in the eyes. Therefore, it is crucial to prioritize providing assistance for removing eye makeup specifically for performers' needs - \textbf{DC1: The prototype should prioritize the eye area to aid in the removal process.}

\paragraph{Invisibility of Makeup Residue to Naked Eyes.} 

When asked about how Xiqu performers determine whether they have completely removed their makeup, all participants (N=15) mentioned relying on a subjective indicator. They either visually inspected the mirror for any remaining makeup with their naked eyes or checked if the towel used to wipe their eyes still had any residual color. (\textit{P8: "After removing my makeup, I would use a facial towel to wipe my face. If there is no color on the towel, it means I have cleaned it thoroughly"}). While these methods can be effective, a portion of the performers (N=10) expressed concerns about not being entirely confident in the thoroughness of their makeup removal. (\textit{P11: "Sometimes I have to wash my face again with facial cleanser multiple times after returning home"}). The standard for judging residual makeup should provide intuitive and objective references. This derived~\textbf{DC2: The prototype should provide an objective and intuitive representation, allowing for easy comparison between the makeup and no-makeup conditions.}

\paragraph{Difficulty of Removing Black \& Pink face paints.}
A unique finding is that despite the different types of operas and roles in China, most performers (N=13) still use pink and black oil-based face paints for their eye makeup. Compared to the flesh-colored face paints used for the base makeup, they are more concerned about removing the pink and black face paints. (\textit{P3: "The oil-based face paints on the face can be easily wiped off, but the black and pink face paints tend to smudge together because of their dark colors"}). Additionally, due to the similarity between black face paints and the color of eyelashes and the difficulty of keeping the eyes open during makeup removal, performers have to observe the black face paints on their eyes multiple times. (\textit{P4: "Makeup removers can easily get into my eyes, and it's challenging to see the remaining black face paints clearly"}). Therefore, the visual representation of residual makeup should offer multi-modal color visualization options -~\textbf{DC3: The prototype should be multi-modal to augment black and pink face paints.}

\subsubsection{Makeup Removal Condition}
\paragraph{Constraints in Space and Time.}
According to most participants reported (N=12), performers usually remove their makeup at the theater. Since performers often go to various venues and not necessarily in a fixed theater, performers need to bring their own makeup and removal tools to different locations, which require portable tools. Additionally, the makeup removal space provided by theaters is sometimes limited, then performers need to queue for their turn to use the designated area. Consequently, they have a restricted amount of time for makeup removal after the performance ends. (\textit{P11: "Some theaters have insufficient makeup removal facilities, therefore, we need to queue up and do it quickly"}). Therefore, the design of the prototype should take into consideration the practical constraints of the makeup removal space and time available -~\textbf{DC4: The prototype should be portable, low-cost, and user-friendly to accommodate the limited space and time available for makeup removal.}

\paragraph{Limited Lighting Conditions in Theatres.}
The backstage conditions at the theater are often unstable, as some performers mentioned that they frequently lack adequate conditions to remove their makeup due to poor lighting in the backstage area. More specifically, the lighting is dim, making it difficult for them to see the actual situation with the naked eye (N=9), (\textit{P15: "Some theater backstage areas have poor conditions with dim lights, making it hard to see subtle remnants of makeup on the face"}). Therefore, the design of the prototype should provide a stable external lighting environment for the performers, free from the constraints of the actual lighting conditions -~\textbf{DC5: The lighting conditions should be stable and optimized to ensure effective makeup removal visualization.}

\subsubsection{Insufficient Awareness regarding Facial Painting's Harmfulness}
Xiqu performers often have demanding schedules and need to take care of their physical well-being. Most of our interviewees (N=12) have reported that they have experienced skin problems or discomfort due to performing since they are too exhausted to remove makeup promptly, leading to unnecessarily long periods of wearing makeup, which in turn triggers various skin issues. \rv{(\textit{P8: "Sometimes I'm so tired that I don't realize I've been wearing makeup for too long. If I become aware that I've been wearing makeup excessively for several days, I'll definitely make sure to remove it promptly."}).} By visualizing the trend of time spent wearing makeup, performers can become more aware of how long they have their makeup on. This awareness aids in timely makeup removal, better skin health management, and prevention of issues like irritation or clogged pores - \textbf{DC6: The prototype should \st{consider providing the visualization of trend in makeup}  \rv{provide some indication of} wearing duration to enhance user awareness and interaction.}



\section{Prototype Design}
With the guidance of six design considerations, we proposed \textit{EyeVis}, an interactive makeup residue visualization system for Xiqu performers. \textit{EyeVis} contains two parts: (1) a fill light \& camera lens magnifier \& eye shields for capturing required images and (2) a mobile app for visualization and user control. We provide the design details and alignments with design considerations in this section.

\subsection{Structural Design}

\begin{figure*}[tbh!]
    \centering
    \includegraphics[width=0.8\textwidth]{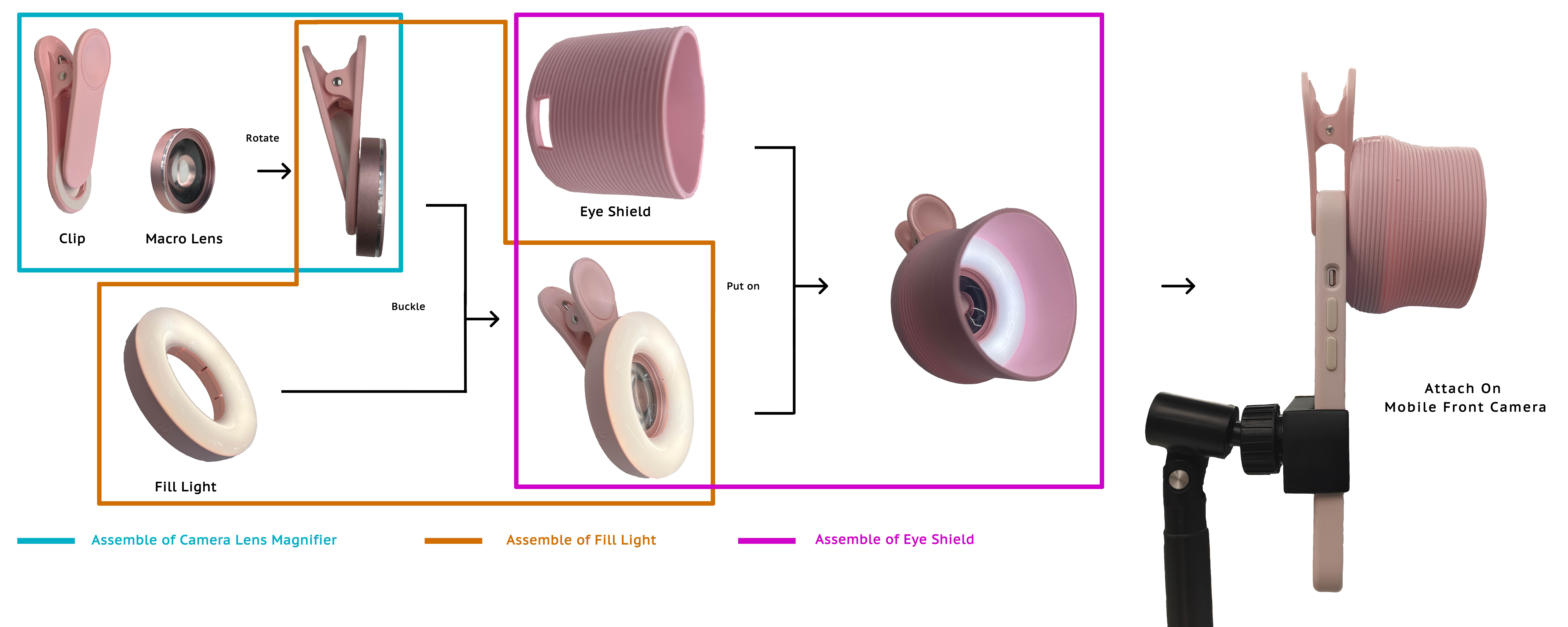}
    \caption{Structural Prototype Assemble: Camera Lens Magnifier (in cyan), Fill Light (in orange), and Eye Shields (in purple)}
    \label{fig: lens}
    \Description{Figure 2 displays the structure prototype assembled, which contains three parts: (1) Camera Lens Magnifier (in cyan), (2) Fill Light (in orange), and (3) Eye Shields (in purple). For the camera lens magnifier, a macro lens is rotated into a clip that can attached to a mobile phone camera. Then, a circular shape fill light is buckled on the camera lens magnifier, and finally, an eye shield is put on the fill light to complete the assembling process.}
\end{figure*}

To design a portable, low-cost, and easy-of-use prototype (\textbf{DC4}), we consider utilizing the camera of mobile phones with low-cost adjustment to match the requirements of capturing images (shown in figure~\ref{fig: lens}). One of the best devices to capture the eye area is the eye tracker~\cite{funke2016eye}, however, it is not applicable when removing makeup because the whole face needs to be cleaned. There are also some existing interactive applications (e.g. \textit{VISIA}~\cite{goldsberry2014visia} and \textit{You Look Good Today}~\cite{Ap3:online}) that capture the whole face for skin imaging analysis, however, systems like \textit{VISIA} have a huge hardware structure that is too cumbersome for the user to move as a portable device. Although systems similar to \textit{You Look Good Today} utilize mobile cameras for skin analysis, which is portable, the performance of the algorithms highly depends on the illumination conditions and the image quality. Therefore, to focus on the eye area instead of the whole face (\textbf{DC1}), we consider utilizing a mobile camera lens magnifier (cyan area in Figure~\ref{fig: lens}) to make sure the images that mobile phones take to satisfy a clear range of focal lengths for mainstream mobile models, and we use a fill light (orange area in Figure~\ref{fig: lens}) with eye shields (purple area in Figure~\ref{fig: lens}) so that users can place eye shields against the skin and block out external light sources, relying only on the fill light source to take the picture, which ensures a constant illumination effect (\textbf{DC5}).


\subsubsection{Camera Lens Magnifier}

Based on formative study results, we aim to design a system that the camera can capture around the eye area (\textbf{DC1}) instead of the whole face. However, the focal length of the camera varies greatly from one mobile phone model to another. In addition, if the full-face shooting mode is used and the eye portion is positioned, on the one hand, it is necessary to deal with the special case of, for example, taking a blurred picture, and on the other hand, the clarity of the cropped picture is not guaranteed, which may lead to a decrease in the accuracy of the image processing algorithm. Additionally, although some mobile devices have a macro lens, it is usually applied in the back, instead of the front. Therefore, we designed a camera lens magnifier, which physically changes the mobile camera lens by applying a clip on the front camera of the mobile phone. We choose to apply a lanthanide optical glass macro lens~\cite{alekseev2021multicomponent} with a 3x effect to achieve the magnification without losing image qualities. The clip is made of ABS plastic. The clip and macro lens can be assembled together by rotating the circle (cyan area in figure~\ref{fig: lens}). With the camera lens magnifier, users can take higher quality pictures from a closer distance than with a mobile phone lens only.

\subsubsection{Fill Light}

To achieve a stable illumination condition (\textbf{DC5}) when taking photos, we believe that blocking external light sources and generating light spontaneously is the most appropriate approach. For light generation, we use a rounded fill light that can be buckled by the camera lens magnifier (orange area in figure~\ref{fig: lens}). The advantage of the circular design is that we allow the built-in lamp beads to be evenly spaced inside the circle, resulting in a stable and well-positioned light source. Since it takes energy for the light beads to emit light, to enhance portability (\textbf{DC4}), we also set up a battery inside so that the fill light can be recharged and used.


\subsubsection{Eye Shields} The eye shields are designed for two purposes: (1) blocking the environment light for achieving stable illumination (\textbf{DC5}) and (2) making sure the images are taken at a constant distance (\textbf{DC1 \& DC2}). We utilized a skin-friendly material, silicone~\cite{berry2007assessing, kao2016duoskin}, to build the eye shields as a circular shape and it can be put on the fill light (purple area in figure~\ref{fig: lens}). Based on the camera lens with magnifier, we calculated a standard focus length as the height of the eye shields, and we set the diameter of the eye shields as 7 centimeters, which suits both fill light and human eyes.


\subsection{Mobile App \& Algorithm Design}
\begin{figure*}[tbh!]
    \centering
    \includegraphics[width=0.8\textwidth]{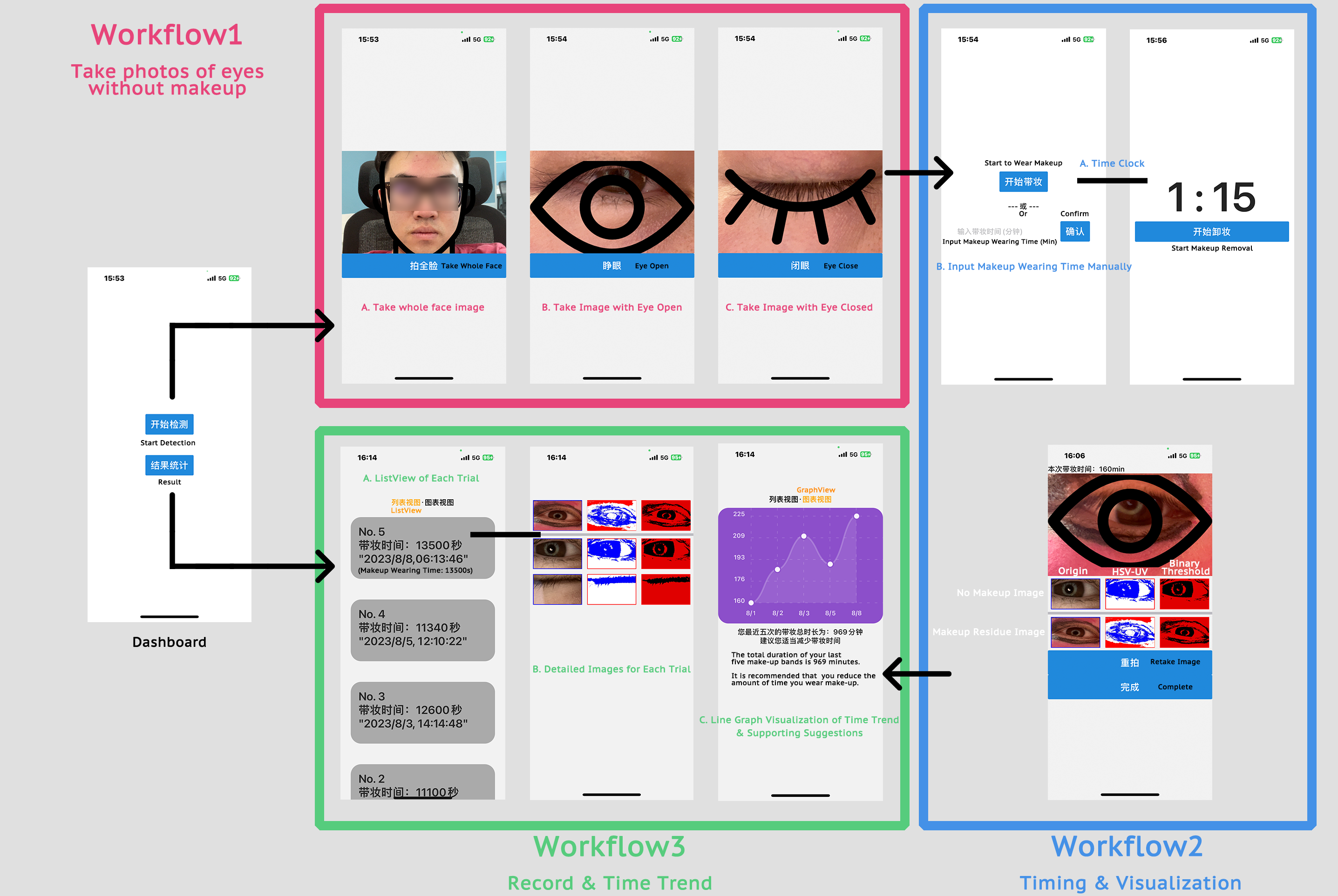}
    \caption{Mobile App Workflow: (1) Take photos of eyes without makeup (whole face, open and close eyes), (2) Timing and visualization (HSV-UV simulation and Binary threshold, comparison w / o makeup), (3) Time trend and record display.}
    \label{fig: app}
    \Description{Figure 3 displays the workflow of the mobile app. Workflow 1: Take photos of the whole face, and eyes without makeup with eyes open and closed. Workflow 2: Record the makeup-wearing time, and take the eye image to see makeup residue by our visualization methods (HSV-UV visualization and Binary Threshold). Workflow 3: Display records and makeup-wearing time trends.}
\end{figure*}

As we utilized the mobile app's camera as a part of our system, it is natural and convenient for users to interact with a mobile app. We designed 3 workflows for Xiqu performers to help them to visualize their makeup residue (\textbf{DC2 \& DC3}) and record their makeup-wearing time (\textbf{DC6}), which are (1) take photos of eyes without makeup, (2) timing \& visualization, and (3) time trend. The details are shown in Figure~\ref{fig: app}. For the usage, users follow these steps: (1) take eye photos without makeup as a baseline, (2) record the makeup wearing time before removing makeup, (3) remove their makeup, (4) take a photo again to visualize makeup residue, and (5) see the usage record and makeup wearing time trend. 

We also embedded three vision-based algorithms inside the app, which are (1) eye feature point localization, (2) hsv-uv simulation, and (3) binary threshold, to achieve the visualization effects of our system (shown in Figure~\ref{fig: io}).

\subsubsection{Prerequisite: Eye feature points localization}

\begin{figure}[tbh!]
    \centering
    \includegraphics[width=0.8\textwidth]{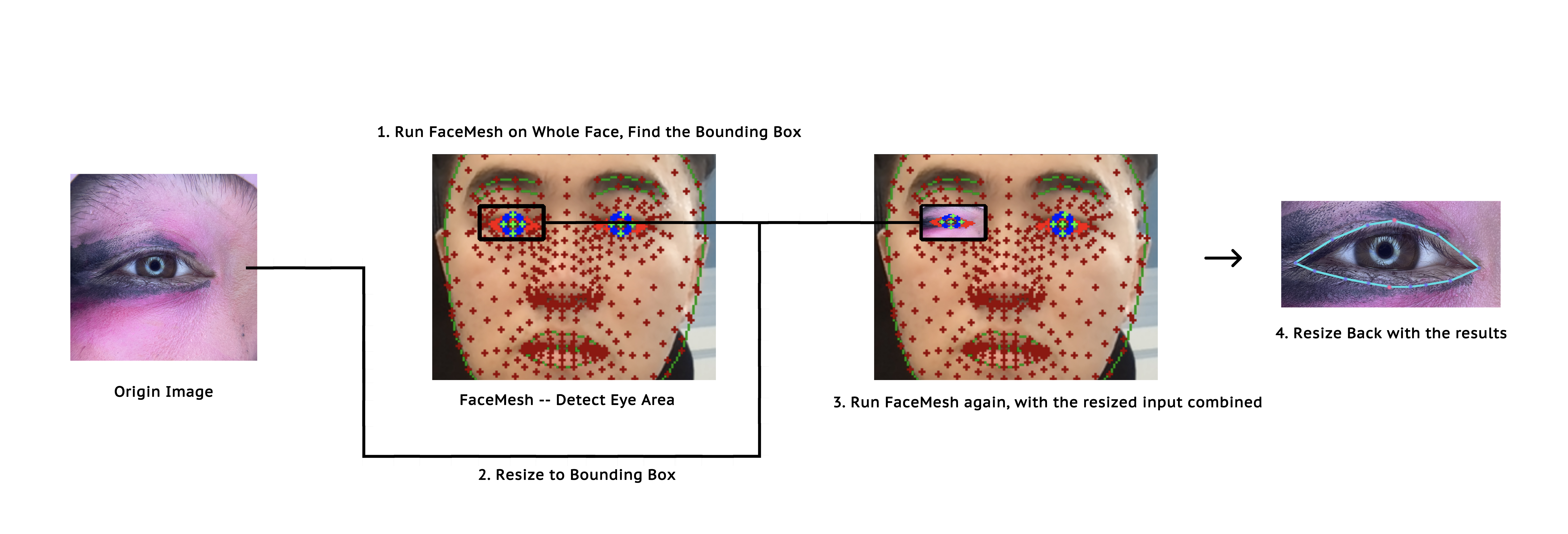}
    \caption{Sample pipeline of Eye feature points localization. \rv{(1) Run Facemesh on whole face as a reference to find the bounding box of eye feature points, (2) Resize the origin image to the bounding box size and replace it with the eye area of the whole face, (3) Run Facemesh again with the input combined image, (4) Extract the eye feature points and resize back to the final result.}}
    \label{fig: prerequisite}
    \Description{Figure 4 displays a sample pipeline of Eye feature point localization. Firstly, based on the whole face picture taken, we run FaceMesh to get the eye feature points localization (a bounding box), then we resize the eye image taken into the bounding box, then we replace the bounding box with the resized eye image to form a combined image, then we run FaceMesh again on the combined image, to get the eye feature points localization on origin eye image.}
\end{figure}

Even though we considered a circular design that fits close to the eyes when designing the structure, there are still cases where the eyes are not centered as each user's face shape is not exactly the same. And the size of the picture taken will be different due to different mobile phone models. To unify this situation, we developed an algorithm of eye feature point localization for each image and resized it to the same size. There are some existing works related to facial feature point localization that achieve high accuracy~\cite{liu2023face, 10.1145/3473682.3480283, xiong2022face2statistics}, however, none of them focuses on the eye area only. To enhance the portability of the deployment, we referred to the face feature points localization and iris landmark detection methods (FaceMesh)~\cite{vakunov2020mediapipe} on Google MediaPipe~\cite{lugaresi2019hadon}. Based on their pre-trained model, we can detect 468 facial landmarks with particular regions (we use a sample human face image, which is provided by the user, to run this process). Next, we collected two index arrays with 16 elements, which represent the left \& right eye areas, and found the bounding box. Then, we replaced the bounding areas with our image input and run the FaceMesh process again. Finally, we can get the localization feature points and map to the original input (setting bounding boxes and paddings). The pipeline details are shown in Figure~\ref{fig: prerequisite}. 

\subsubsection{Workflow1: Take photos of eyes without makeup}

To let users compare w/o (with / without) makeup images (\textbf{DC2}), we designed the workflow of letting users take pictures of no makeup first. In this step, if the user first opens the app, then the user needs to take a photo of the whole face as input for the prerequisite algorithms. Next, the user needs to take two pictures, one with eyes open and one with eyes closed (shown in Figure~\ref{fig: app}, Workflow1), so that when taking subsequent pictures with makeup if the user doesn't pay attention to the eyes open or closed, the user will still have a similar picture without make-up to match for better comparison. The matching scheme is to calculate the distance of the highest / lowest eye feature points to determine the degree of opening and closing of the eyes.

\subsubsection{Workflow2: Timing \& Visualization}

\paragraph{Timing} After taking photos without makeup, the user starts to wear the makeup with the face paint. To record the makeup-wearing time (\textbf{DC6}), we designed a timing scheme. For ease of use, the user can choose to record the wearing time by a time clock or input the wearing time manually (shown in Figure~\ref{fig: app}, Workflow2, A \& B).

\begin{figure}[tbh!]
    \centering
    \includegraphics[width=0.8\textwidth]{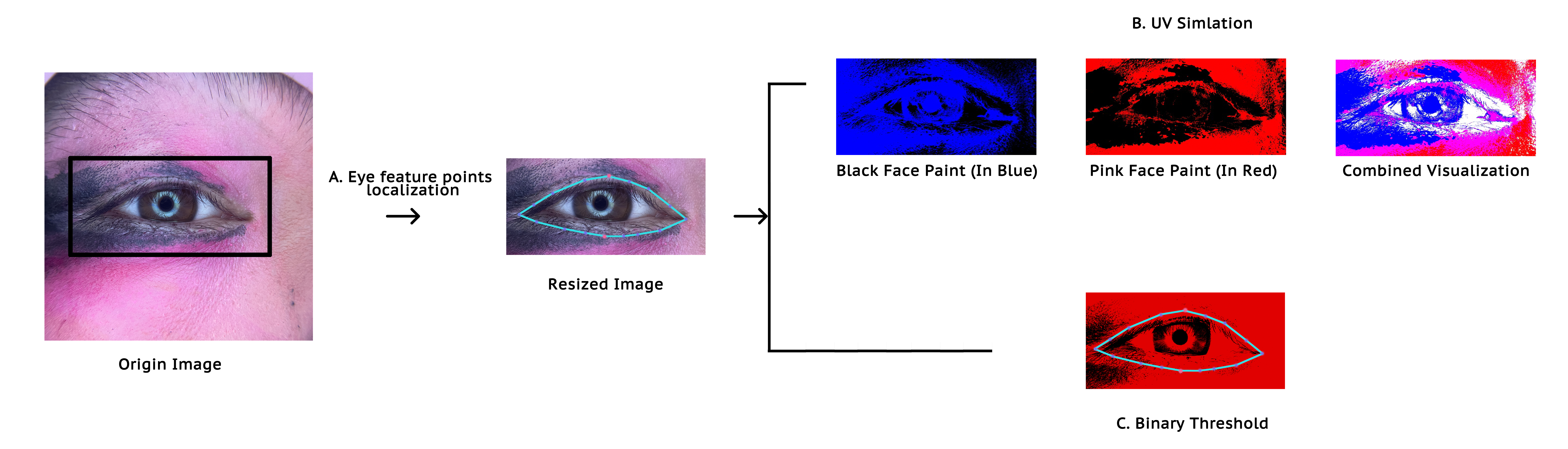}
    \caption{Overview of Vision-based Algorithms: firstly run (A) eye feature points localization (clip and resize the image based on the bounding box), then run (B) HSV-UV simulation (we provide blue color for black face paint, and red color for pink face paint visualization) and (C) Binary Threshold to visualize the makeup residual.}
    \label{fig: io}
    \Description{Figure 5 displays the algorithm input and output pipeline. From the origin image taken by our system, we run Eye feature points localization algorithm to get the resized image, then we run the Visualization algorithms of HSV-UV simulation (we provide blue color for black face paint, and red color for pink face paint visualization) and Binary Threshold.}
\end{figure}

\paragraph{Visualization: HSV-UV Simulation} To augment the black and pink areas of residual face paint (\textbf{DC3}), we first tried special light sources~\cite{maverakis2010light}. Ultra-Violet (UV) light has been widely used in skin disease diagnosis, and it is also effective in allowing cosmetic residues to fluoresce~\cite{bispo2021ultraviolet}. However, the eye area can be irreversibly and severely damaged by exposure to UV light~\cite{matsumura2004toxic}. Therefore, we try to make the UV simulation by utilizing color spaces and image processing \& filtering. Although our system is designed for stable illumination, the RGB color space is sensitive to very small changes in brightness (R, G, B values can vary greatly). Therefore, we utilized HSV color space to filter similar colors of black and pink face paints. HSV color filtering contains these steps: (1) enhance the intensity of the blue channel, (2) change from RGB to HSV color space, then set HSV thresholds (lower and upper bound) of black and pink face paints, (3) apply HSV filtering and transfer to single-color images (blue for black face paints, and red for pink face paints). Figure~\ref{fig: io}, B, shows a sample of the algorithm effect, the blue image means black face paints are detected, while the red image links with pink face paint. The pink color in the combined image shows the intersection of black and pink face paint.

\paragraph{Visualization: Binary Threshold} To better describe the situation around the eyes, we designed another image-processing algorithm to visualize the edges \& contours of the eye-around area. The idea is to transfer the color space of the image from RGB to gray, then we set a lower bound and upper bound for thresholding: pixels inside the bound will be marked as black, otherwise white. Finally, we apply edge detection and give the output image a striking color map to replace the white area. Figure~\ref{fig: io}, C,  shows a sample of the algorithm effect, the user can see edges \& contours of eyeliner around the eye area.

Finally, there are two rows of image sets, which represent the w/o makeup images. For each row, the original image, HSV-UV simulation image, and binary threshold image are shown as a set for the user to check and compare.

\subsubsection{Workflow3: Time Trend} After recording makeup wearing time for each time, the user can see a time trend visualization (\textbf{DC6}) provided by \textit{EyeVis}. We provided 2 types of record visualization: (1) a list view of each use, the user can click into each use for rechecking the images captured for each use, and (2) a line graph to represent the time trend of the last five makeup experiences (shown in Figure~\ref{fig: app}, Workflow 3, A \& C).

\subsection{Implementation}

\subsubsection{Algorithm Development}

To localize the eye feature points, we used the MediaPipe~\cite{lugaresi2019hadon} Python library (version 0.9.1, which supports face landmark detection) and Numpy (version 1.24.3) for image cropping and resizing. We also utilized the OpenCV~\cite{bradski2000opencv} Python library (version 4.8.0.74) for HSV-UV simulation and Binary Threshold visualization. Finally, we integrated all code scripts into the back end of the mobile app.

\subsubsection{Mobile App Development}

To make our app accessible to more users, we use a cross-end framework, react-native~\cite{eisenman2015learning}~\footnote{https://reactnative.dev/}, to develop our mobile app in JavaScript scope. For algorithm deployment, we developed a back-end by using FastAPI of Python~\footnote{https://fastapi.tiangolo.com/} and deployed the script to a Tencent Cloud~\footnote{https://www.tencentcloud.com/} remote server (running on Linux, CentOS 7.9). Then, we deploy our code to both XCode and Android Studio for iOS and Android application packages. Finally, we release our system by using an online app distribution service called Pgyer~\footnote{https://www.pgyer.com/}.
\section{Technical Evaluation}

To evaluate the robustness of the algorithms, we used \textit{EyeVis} to collect a testing database that contains 259 eye images from four co-authors. The database collection procedures include: (1) take eye pictures with no makeup, (2) take eye pictures with full makeup (with pink and black face paint), and (3) take eye pictures with makeup residue \rv{(i.e. remove the face paints in parts of the eye-around area but not all). To extend the generalizability of the dataset, we collected images that contain both open and closed eyes, and the number of images in different categories is evenly distributed.} Although there are many large-scale models that are gradually being applied to image segmentation~\cite{kirillov2023segment, liu2023grounding}, their effectiveness is limited to recognizing more common objects, and they are generally ineffective for images with only eyes and makeup. Therefore, all images are manually labeled with makeup areas (black \& pink face paints and eye-around contours). Technically, we use Computer Vision Annotation Tool (CVAT~\cite{cvat_ai_corporation_2023_8239758}) to manually label 3 classes (pink face paint, black face paint, eye) with polygons for each image.

\subsection{Illumination Stability}

\begin{figure}[tbh!]
    \centering
    \includegraphics[width=0.8\textwidth]{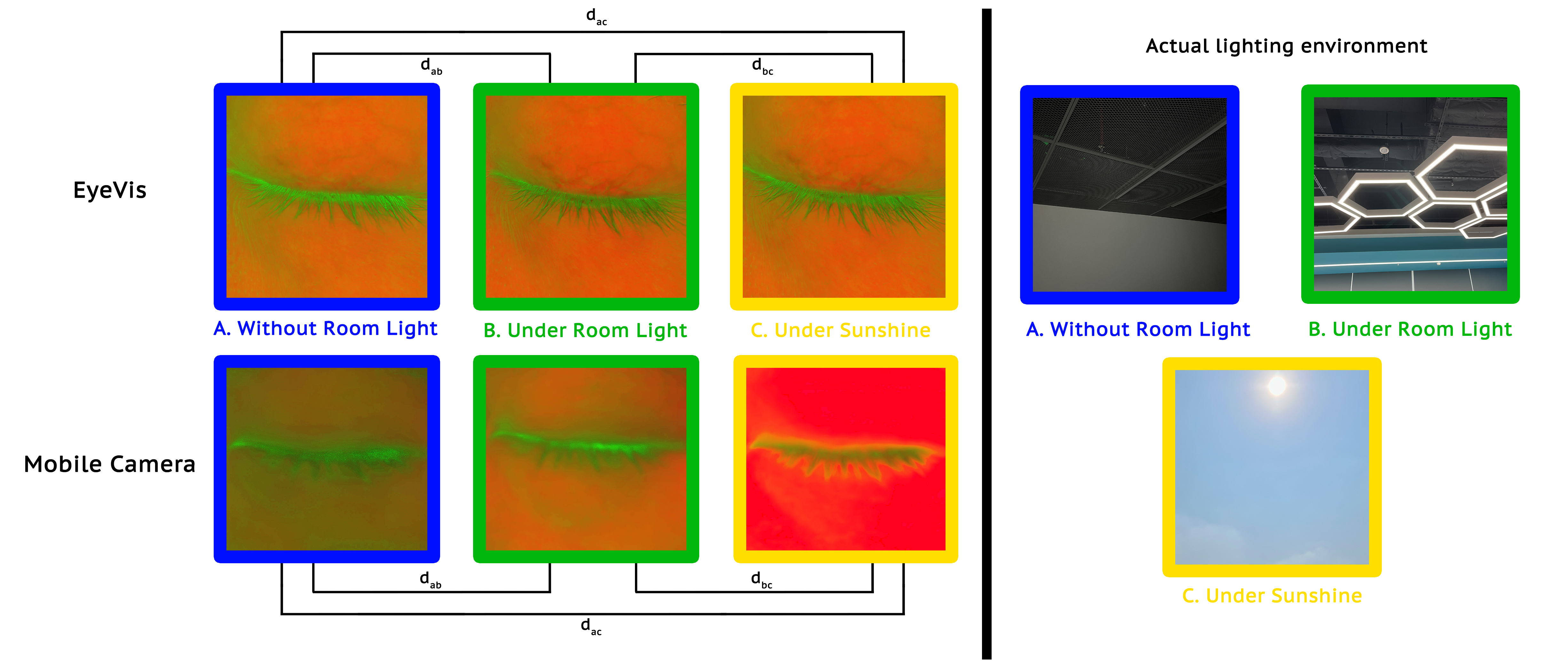}
    \caption{\textbf{LEFT}: Sample Illumination Test: w / o EyeVis in HSV color space. \rv{\textbf{RIGHT}: Actual lighting surrounding environments. } The images are taken under 3 different lighting conditions: a) without room light, b) under room light, and c) under sunshine.}
    \label{fig: illumination}
    \Description{Figure 6 displays sample illumination test results that contain 6 subfigures (2 rows * 3 columns). The first row displays results taken by \textit{EyeVis}, and the second row displays results taken by mobile camera only. For three columns, it claims the image taken by various lighting conditions: (1) without room light, (2) under room light, and (3) under sunshine. The actual lighting environment of these 3 condition are also provided.}
\end{figure}

To validate the stability of the illumination of \textit{EyeVis}, we tested 10 groups of data, each group contains a series of images that are taken by the same eye \rv{with the same camera settings ($ISO = 200$, $Shutter = \frac{1}{50}s$, $Focal \ Length(f) = 26mm$, $Aperture = \frac{f}{2.2}$, $White \ Balance = 5200K$)}, in various lighting surrounding conditions: (1) without room light, (2) under room light, and (3) under sunshine. Then we transformed the images in HSV color space, and we calculated the Euclidean distance~\cite{danielsson1980euclidean} ($d = \sqrt{{d_h}^2 + {d_s}^2 + {d_v}^2}$) of each two images ($img_0$, $img_1$) in each pixel in terms of hue ($d_h \ = \ min(abs(h_1 \ - h_0),\ 360-abs(h_1 \ - h_0)) / 180 $), saturation ($d_s \ = \ abs(s_1-s_0)$), and value ($d_v \ = \ abs(v_1-v_0)/255$). Figure~\ref{fig: illumination} shows a sample of the grouped data, compared with images taken by a mobile phone's front camera directly at the same distance as using \textit{EyeVis} (we used iPhone 13's camera for testing). Table~\ref{table: illumination} shows our test average results, based on the average data, we observed that all Euclidean distances of the two chosen images are significantly lower than taking images by mobile directly. Therefore, we can conclude our system achieves a stable illumination environment.

\begin{table}[tbh!]
\centering
\begin{tabular}{|c|c|c|c|}
\hline
                       & Avg $d_{ab}$ & Avg $d_{ac}$  & Avg $d_{bc}$ \\ \hline
EyeVis & 0.86         & 0.73        & 0.69        \\ \hline
Mobile Camera          & 1.91        & 4.60        & 3.09        \\ \hline
\end{tabular}
\caption{Average Euclidean Distance Comparison Under Different Lighting Conditions: w / o EyeVis}
\label{table: illumination}
\Description{Table 1 shows the Euclidean distance of two images in HSV color space (without room light versus under room light -- 0.86: 1.91, without room light versus under sunshine -- 0.73: 4.60, and under room light versus under sunshine -- 0.69: 3.09) comparing EyeVis and mobile camera.}
\end{table}

\subsection{Visualization Methods}

We evaluate the robustness of visualization methods: 1) eye feature points localization, 2) hsv-uv simulation, and 3) binary threshold, by comparing the regional overlap between algorithm results and manually labeled area. We utilized the Shapely~\cite{gillies2013shapely} Python library to do the polygon-area transformation and overlap calculation. 

\begin{figure*}[tbh!]
    \centering
    \includegraphics[width=0.9\textwidth]{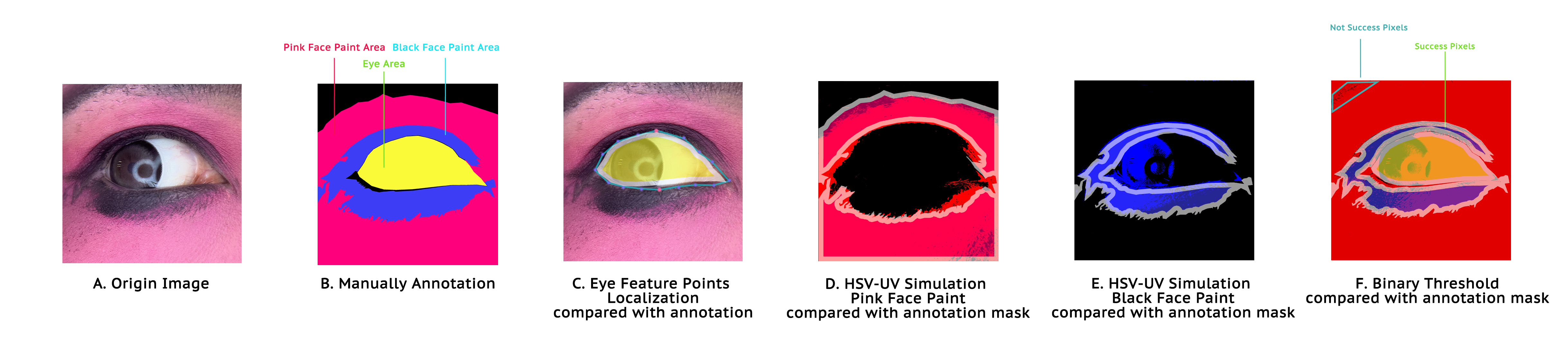}
    \caption{Sample Technical Evaluation of Visualization Methods: A shows a sample origin image; B shows the related manually annotated image; C, D, and E show the overlap between algorithm results and manually annotated area, in terms of the eye area, pink face paint area, and black face paint area, respectively; F shows ``success'' points (inside black area) in the binary threshold image.}
    \label{fig: vis-tech}
    \Description{Figure 7 displays a sample technical evaluation of visualization methods, which contains 6 subfigures (A - F). A shows the origin image taken by EyeVis, B shows the image with manual annotation, which contains eye area, pink face paint area and black face paint area. C shows the eye feature points localization algorithm results, compared with eye annotation. D and E show the HSV-UV simulation result that contains pink and black face paint compared with the manual annotation mask. F shows the "success" and "no success" points in the binary threshold image.}
\end{figure*}

\subsubsection{Eye Feature Points Localization}

To evaluate the accuracy of eye feature point localization, we used our manually marked eye area to compare with the algorithm results, and we define $a_1$ = \textit{algorithm results inside manually marked eye area}, $a_2$ = \textit{manually marked eye area}, $r_{eye} = \frac{a_1}{a_2}$ as the quantitative value that indicates the accuracy (see Figure~\ref{fig: vis-tech}, C, for the region comparison). The areas of the algorithm results are formed by polylines of the 16 feature points.

\subsubsection{HSV-UV Simulation}

For the evaluation of hsv-uv simulation, we also define $a_3$ = \textit{algorithm results inside manually marked pink area}, $a_4$ = \textit{manually marked pink area}, $r_{pink} \ = \frac{a_3}{a_4}$ and  $a_5$ = \textit{algorithm results  inside manually marked black area}, $a_6$ = \textit{manually marked black area}, $r_{black} \ = \frac{a_5}{a_6}$ as quantitative indicators to justify the accuracy (see Figure~\ref{fig: vis-tech}, D \& E). 

\subsubsection{Binary Threshold}
To test the robustness of binary threshold effects, we examined the area outside of the manually delineated eye range, and if the pixel labeled black in the threshold range happened to fall within the manually delineated black paint area, the pixel would be labeled as "success". We define $n_1$ = \textit{number of inside-threshold pixels marked as ``success"} (outside-threshold pixels marked as ``not success''), $n_2$ = \textit{number of inside-threshold pixels outside eye area}, $r_{bin} \ = \frac{n_1}{n_2}$ as the quantitative indicator to justify the accuracy of the binary threshold (see Figure~\ref{fig: vis-tech}, F).

\begin{table}[tbh!]
\centering
\begin{tabular}{|l|l|l|l|l|}
\hline
                 & $r_{eye}$ & $r_{pink}$ & $r_{black}$ & $r_{bin}$ \\ \hline
Avg Overlap Rate & 96.1\% & 80.4\%  & 85.2\%   & 82.8\% \\ \hline
\end{tabular}
\caption{Average Overlap Rate among eye feature points localization ($r_{eye}$), HSV-UV simulation ($r_{pink}$ and $r_{black}$), and Binary Threshold ($r_{bin}$).}
\label{table: binary}
\Description{Table 2 shows the average overlap rate among eye feature points localization (96.1\%), HSV_UV Simulation of pink (80.4\%) and black (85.2\%) area, and Binary Threshold(82.8\%).}
\end{table}

Table~\ref{table: binary} shows the testing results among 259 images in our self-maintained dataset. $r_{eyes}$ achieves the highest average overlap rate of 96.1\%, which is a testament to the robustness of the algorithm on the task of locating the eyes. For the HSV-UV simulation method, the results for the pink area ($r_{pink}$) are slightly lower than the black area ($r_{black}$), and we speculate that this may be due to the fact that each person applies oil colors in different shades, and thus the HSV may not have covered all the cases when setting up the color filter. For $r_{bin}$, some images are affected by things other than the eye, such as eyebrows and other objects that resemble a black border, potentially causing some pixels to be ``not successful''. However, the average overlap rate for each indicator still remained above 80\%. As we developed our algorithms without machine learning to reduce the burden of back-end computing, such an overlap rate indicates our algorithms can achieve considerable effects, and are ready to be deployed and tested by users.

\section{Deployment Study}

To examine the usability and effectiveness of \textit{EyeVis}, we ran a 7-day deployment study with 12 participants. Based on the study results, we discussed and explored potential design implications for future work. To our knowledge, it is the first time a prototype of this kind has been created and tested in a long-term deployment study.

\subsection{Participants \& Apparatus}
We recruited 12 participants (3 males, and 9 females, aged 16-55) by posting recruitment messages on the online community. All participants are from Xiqu troupes or academies and have experience in Xiqu makeup. \rv{Before the experiment, each participant signed a consent form (approved by the university's ethical review) to agree to collect user data.} As we deployed our back-end in a remote server, and participants can use the prototype anywhere on any Android or iOS device. For remote users, we also sent a package of fill lights, camera lens magnifier, and eye shields to each of them. \rv{100 RMB was given to each participant as the reimbursement.}

\subsection{Task Design \& Procedure}
\label{sec: task design -- procedure}
We designed a 7-day user study to explore the usability and effectiveness of our system. \rv{Before the study officially starts, we record eye images (by using the same camera settings as \textit{EyeVis} without letting participants see the visualization and remove accordingly) in a daily makeup removal process for each participant, so that we can perform the baseline (daily makeup removal) \& \textit{EyeVis} comparison. } For Day 1, we have an introduction session for kicking off, we introduce users to (1) how the structure part is installed, (2) how the software part is installed, and 3) how to use the app for the makeup removal experience. Then, participants went through three workflows of the mobile app. Before taking photos without makeup, participants were asked to clean their faces carefully, to avoid residue before applying make-up. Then participants wear eye makeup and remove it for a test. Participants were also requested to complete a NASA TLX form~\cite{hart1988development} to document their daily makeup removal routine experience, serving as a baseline data recording. After the introduction session, for Day 2 to Day 7, participants were asked to repeat the makeup and makeup removal process at least 4 more times (i.e. a total of at least 5 times of using) during performance or daily practice, to examine the effects of makeup-wearing time trend. To simulate the most realistic use, participants were given the opportunity to use their personal favorite make-up removers to remove their make-up. For each use, we recorded the photos that participants took and the makeup removal time in the back end for analysis, and the consent of the participants was obtained. Finally, on Day 7, participants were asked to fill out a NASA TLX form \rv{and a 7-point Likert scale questionnaire} to evaluate the usability, comfortability, and effectiveness of the system~\cite{hornbaek2006current}, and a final semi-structured interview was conducted.

\subsection{Data Analysis}

Four co-authors open-coded the deployment study data in terms of qualitative statistics (user feedback in semi-structured interviews, \rv{and ratings in the NASA TLX forms, in terms of the usability of the system, and ratings in 7-point Likert scale questionnaires regarding the features of the system}) and quantitative statistics (\rv{the images contained by using \textit{EyeVis} and daily makeup removal}). Then we summarized the data in terms of the system's usability, the considerable long-term effects of use, \rv{and the quantitative comparison between using \textit{EyeVis} and \textit{Daily Makeup Removal}}.

For quantitative comparison of makeup residue, we take the following approaches: 

\begin{enumerate}
    \item  Run the \textit{EyeVis} back-end algorithm once with the captured daily makeup removal image as input, we can get the area where the pink face paint and black face paint are located, and calculate the ratio of the area it occupies to the whole image area ($r_{p\_baseline}=\frac{pink \ face \ paint \ area}{whole \ baseline \ image \ area}$, $r_{b\_baseline}=\frac{black \ face \ paint \ area}{whole \ baseline \ image \ area} $).
    \item We can get the area of pink and black face paint in the eye image captured by the user every time the participant uses \textit{EyeVis}, and calculate the ratio of the area to the whole image ($r_{p\_EyeVis} \ = \ \frac{pink \ face \ paint \ area}{whole \ EyeVis \ image \ area}$, $r_{b\_EyeVis} \ = \ \frac{black \ face \ paint \ area}{whole \ EyeVis \ image \ area} $). 
    \item Get the mean value of the data ($r_{p\_EyeVis\_mean} \ = \ \frac{r_{p\_EyeVis}}{5}$, $r_{b\_EyeVis\_mean} \ = \ \frac{r_{p\_EyeVis}}{5}$) when the user uses the data five times and compare this mean value with EyeVis.
\end{enumerate}

Note that we don't use the makeup removal time as an indicator of how effective the system is compared to daily makeup removal, because when using the \textit{EyeVis} system, users tend to double-check to see if they're getting it off cleanly, which can lead to longer make-up removal times. Similarly, the time taken to remove makeup depends on the thickness and complexity of the makeup. Therefore, we only use ``how completely the makeup is removed'' to judge the system's effectiveness.


\section{Result}

\subsection{System Usability}
\subsubsection{Overall Usage of the System}
Overall, all participants indicated that the usage of an assistant makeup removal system was helpful for them to remove makeup more effectively. The system's usability was evaluated by the ratings of the NASA TLX questionnaire. \rv{In terms of performance (makeup removal completeness)}, they agreed that compared with baseline (makeup removal without \textit{EyeVis}, recorded before introducing \textit{EyeVis} to participants) (\textit{Md}=2, \textit{IQR}=2), with the assistance of \textit{EyeVis}, they can remove makeup more completely (\textit{Md}=6, \textit{IQR}=1.25). Participants also appreciated the EyeVis system, noting its significant assistance in reducing the efforts (\textit{Md}=4, \textit{IQR}=1.5) \rv{and frustration (\textit{Md}=3, \textit{IQR}=1.25)} with the makeup removal process. They found that utilizing the assistive system reduced the physical (\textit{Md}=2, \textit{IQR}=2.25), mental (\textit{Md}=3, \textit{IQR}=3) \rv{and temporal (\textit{Md}=3, \textit{IQR}=2) demands} involved in removing their makeup, especially when compared to their daily routine, which often required repeated wiping to ensure that all the makeup was properly removed.~\textit{"With the EyeVis system, I can easily determine when to stop wiping, avoiding over-cleansing and discomfort due to excessive rubbing."-P3.}


\begin{figure*}[tbh!]
    \centering
    \includegraphics[width=\textwidth]{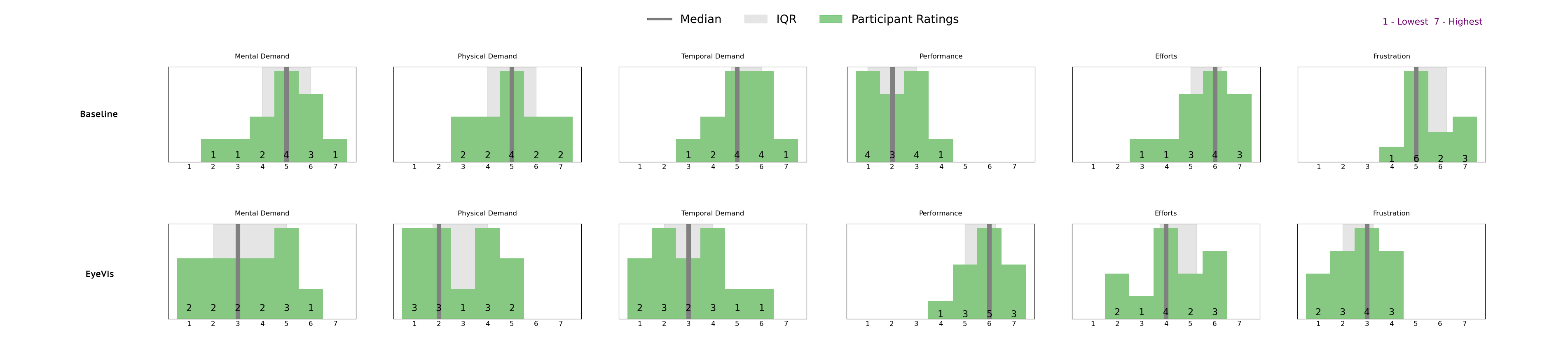}
    \caption{\rv{Participants' ratings (regarding the usability of the system) on a scale of 1-7 using NASA Task Load Index. \textbf{Top: }ratings for daily makeup removal. \textbf{Bottom: }ratings for makeup removal by using \textit{EyeVis}.}}
    \label{fig: nasa-daliy}
    \Description{Figure 8 contains 2 rows (baseline versus using EyeVis), each row displays a NASA task load index questionnaire results in terms of mental demand, physical demand, temporal demand, performance, efforts, and frustration.}
\end{figure*}

\subsubsection{Visualization methods}
\begin{figure*}[tbh!]
    \centering
    \includegraphics[width=\textwidth]{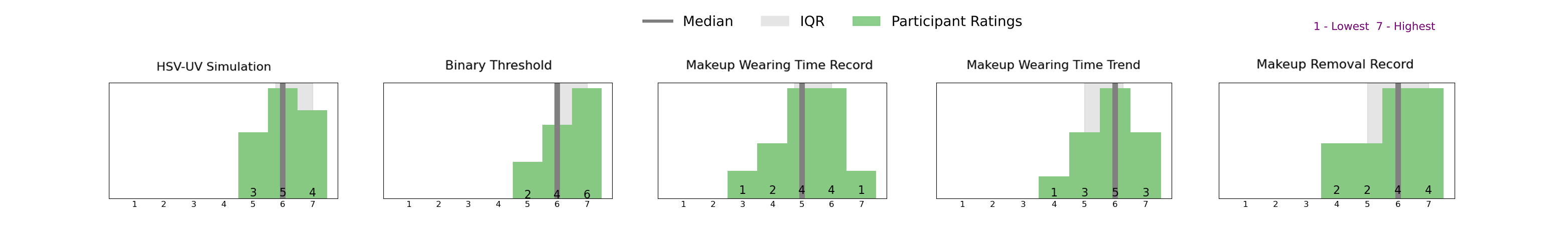}
    \caption{\rv{Participants' ratings (regarding the feature of the system) on a 7-point Likert scale questionnaire. For the performance scale, a higher rating means better performance.}}
    \label{fig: likert-feature}
    \Description{Figure 9 displays a 7-point Likert scale questionnaire results, in terms of ratings of "HSV-UV Simulation", "Binary Threshold", "Makeup Wearing Time Record", "Makeup Wearing Time Trend", and "Makeup Removal Record".}
\end{figure*}

 The ratings of the 7-point Likert scale questionnaire evaluated different visualization methods of the system (shown in Figure~\ref{fig: likert-feature}). \textit{EyeVis} provides users with two visualization methods, HSV-UV Simulation and Binary Threshold, to identify makeup residue. There are no significant advantages or disadvantages between ~\textbf{HSV-UV Simulation} (\textit{Md}=6, \textit{IQR}=1.25) and \textbf{Binary Threshold} (\textit{Md}=6.5, \textit{IQR}=1) based on the ratings. Users, on average, expressed their appreciation for both techniques based on their preferences. However, most participants (N=10) did appreciate specific aspects of Binary Threshold, as P10 stated,~\textit{"I prefer Binary Threshold because it creates a more intuitive color contrast around eye area"}. In contrast, limited participants preferred HSV-UV Simulation (\textit{P1}, \textit{P6}), saying, ~\textit{"I mostly focus on the HSV-UV visualization as I want to quickly identify and remove the black \& pink face paint inside the eye"}. Additionally, users utilized the visualization approach to monitor different aspects of makeup residue, ranging from overall residual makeup to more detailed areas,~\textit{"HSV-UV helps me gauge the extent of general residue, while Binary allows me to examine the residue in more detail"-P7.}


\subsubsection{Objective Feedback of Time Trend}
Participants highly valued the \textit{EyeVis} system's makeup-wearing time record (\textit{Md}=5, \textit{IQR}=1.25) and makeup-wearing time trend (\textit{Md}=6, \textit{IQR}=1.25) as the most crucial aspect of the system. They found the time recording intuitive and user-friendly, \textit{P4} expressed,~\textit{"I simply click on the interface at the start and end of my makeup process, and it automatically records the time. This allows me to keep track of my skin's condition more accurately"}. The makeup-wearing time trend further enhanced participants' awareness. \textit{P7} shared,~\textit{"After wearing makeup for three days, I checked the trend feature and realized that I had been applying makeup too frequently, which could lead to skin problems. Previously, I might have ignored or not remembered this. So, I followed a meticulous skincare routine"}. Some participants also mentioned additional benefits of viewing recordings, which helped them monitor their skin condition more precisely.~\textit{"Occasionally, I compared my eye pictures from a week ago with today's and noticed that my dark circles had worsened. It made me realize I should take some time off" -P10}. We also observed that the time trend is an initial approach in skincare recommendations, as \textit{P7} stated: \textit{"It's nice to be able to visualize how long I can wear makeup, but it would be nicer to recommend skincare products or skincare regimens based on my actual needs."}

\begin{figure*}[tbh!]
    \centering
    \includegraphics[width=0.8\textwidth]{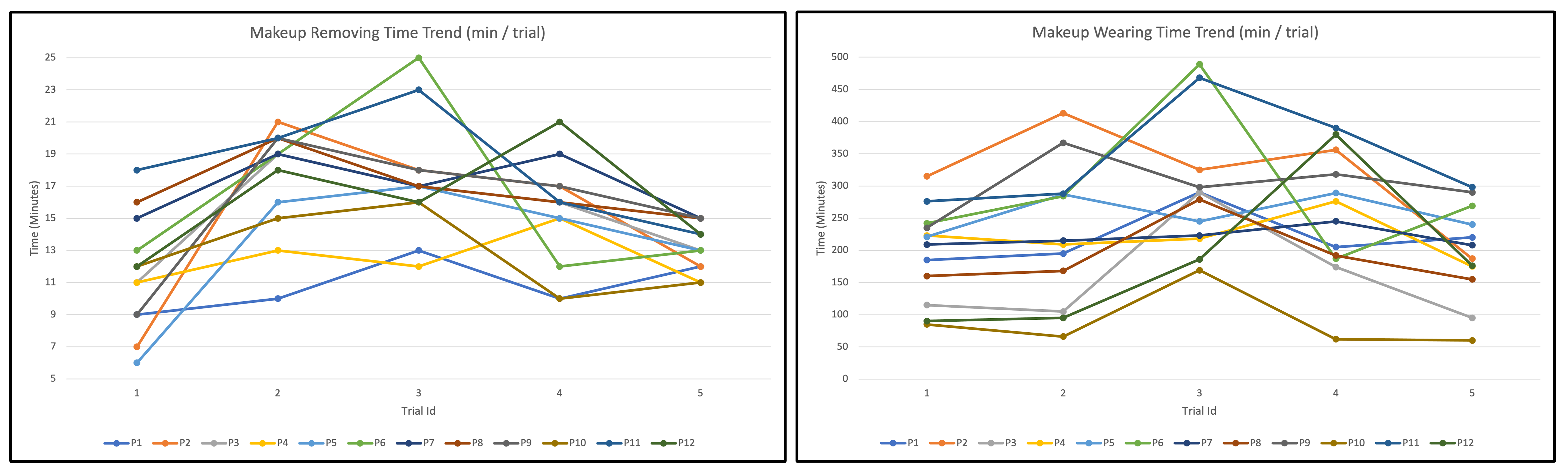}
    \caption{The makeup-wearing \& removing time trend among all participants in five trials. \textbf{LEFT}: Makeup removing time trend. Overall, participants enhanced their makeup removing time to familiarize themselves with specific visualization methods of the system, then it reduced when they got used to it. \textbf{RIGHT}: Makeup wearing time trend. Generally, participants \st{reduced their makeup-wearing time and} enhanced their awareness of makeup removal.}
    \label{fig: makeup-time}
    \Description{Figure 10 displays the makeup-wearing (Right) \& removing (Left) time trend among all participants in the deployment study. Overall, participants improved their makeup removal time by becoming acquainted with the specific visualization techniques of the system, and participants lowered their makeup usage and heightened their level of attentiveness.}
\end{figure*}

\subsection{Persistent Effects}

\subsubsection{Consistent Trend of Makeup Removal Time}
Figure~\ref{fig: makeup-time} shows a general trend in makeup removal time for 12 participants. As each participant has different scenes with make-up (rehearsal or performance) and different make-up removal habits, we can see that the initial values in the graphs are quite different, but we can still find some patterns in them: First, during the initial usage of the system, participants received assistance from the researchers. However, as they continued interacting with the system over time, their interest in other features gradually increased. As a result, the second usage of the system showed a slight upward trend. Participants took a casual approach to makeup removal initially, which resulted in a relatively short time spent removing makeup. Furthermore, as they progressed and understood the importance of removing makeup thoroughly the duration of the procedure trended upwards. As the participants continued to use the assistant system during the makeup removal process, they gradually became more familiar with its functionality and developed a routine. This familiarity and routine led to increased efficiency in removing makeup, resulting in a decrease in the overall duration of the process. As \textit{P5} stated,~\textit{"At the beginning, I wasn't very familiar with the Usage of EyeVis, but after using it three or four times, I gradually mastered the tricks and it allowed me to remove my makeup more quickly"}. Figure~\ref{fig: makeup-time} also shows a positive correlation between the makeup-removing time and makeup-wearing time, especially for \textit{P6} and \textit{P11}. This suggests that participants who wore makeup for longer took more time to remove it completely,~\textit{"During a long-time performance, makeup needs to be touched up repeatedly, resulting in a thicker layer. Consequently, it takes me more time to remove the makeup."-P12}.

\subsubsection{Change in Makeup Removal Practices}

We observed that some participants' makeup removal habits changed after the 7-day experiment, e.g. some participants who had previously paid less attention to the make-up removal process paid more attention to this process: As \textit{P9} stated: \textit{"Previously I was very casual about removing my make-up and after using EyeVis I would subconsciously check to see if it had been removed completely"}. For experienced and skin-conscious actors, their make-up removal habits are not subject to much change. \textit{P2} mentioned that: \textit{"Having been doing Xiqu make-up for many years now, it's a fact that eye make-up is hard to remove. However, since I have more experience, my make-up removal habits have not been affected much."} Additionally, there were some participants who, as heavy users of face paint, had increased the importance they placed on removing their makeup: \textit{"As a frontline actor, I wear my make-up for a very long time every day, and by the time I take it off I'm often tired from several performances, so I remove my make-up in a haphazard manner. This system has really helped me to improve my make-up removal habits and hopefully, my skin will slowly improve." - P6}

\subsection{\rv{Quantitative Comparison of Makeup Residue: \textit{EyeVis} vs \textit{Daily Makeup Removal}}}


\rv{We report each participant's ratios of specified regions (pink \& black face paints) in Appendix~\ref{append: quant}. Based on our collected data: $r_{p\_baseline}$ (\textit{avg} = 19.84\%, \textit{std} = 6.51\%), $r_{p\_EyeVis\_mean}$ (\textit{avg} = 4\%, \textit{std} = 1.66\%), $r_{b\_baseline}$ (\textit{avg} = 19.48\%, \textit{std} = 6.81\%), and $r_{b\_EyeVis\_mean}$ (\textit{avg} = 3.63\%, \textit{std} = 1.61\%), we observe that: (1) \textbf{Significant Reduction in Face Paint Residual Compared with Baseline.} There's a marked decrease in the percentage of both pink and black paint areas when measured using the \textit{EyeVis} system compared to the baseline. This could suggest that the \textit{EyeVis} system is more efficient or conservative in removing the makeup. (2) \textbf{Consistency and Variability.} The \textit{EyeVis} system measured two colors with a small standard deviation, indicating better consistency. This means that the \textit{EyeVis} system standardized the makeup removal process more effectively than the baseline method (daily makeup removal), resulting in more consistent results across participants. (3) \textbf{Diversity in Individual Cases.} Even though the system shows greater consistency, there is still a range of responses among participants. This diversity might be attributed to individual differences, such as variations in face size or shape, which could affect the makeup removal completeness.}
\section{Discussion}
In this section, we summarize the most important findings, observations, and limitations of our research. We also highlight a few design takeaways from the research and potential future directions.

\subsection{Key Contributions}
We investigated the practices and challenges encountered by Chinese traditional opera (Xiqu) performers in relation to skin condition issues, by conducting an online survey and semi-structured interviews. Specifically, we identified that incomplete makeup removal is the key cause due to human factors, and designing an interactive makeup residue visualization application might be helpful for Xiqu performers in aid of their skincare. From there we derived six design considerations in section~\ref{sec: formative-findings}.

Then, we implemented these design considerations in a prototype called \textit{EyeVis} with these parts: mobile app + fill light + camera lens magnifier + eye shields, and in the visualization part, we embedded image processing algorithms including (1) eye feature points localization, (2) hsv-uv simulation, and (3) binary threshold. We conducted a technical evaluation for illumination stability and visualization methods by utilizing a self-maintained eye image database. We also evaluated the usability and long-term effects of \textit{EyeVis} by running a 7-day deployment study with 12 participants. We arrived at the following key conclusions and insights to guide future design: (1) Two Visualisation methods enhance the color display of makeup residue to help users confirm that it has been removed, (2) \textit{EyeVis} is designed to meet the portability requirements of users, (3) Long-term use of \textit{EyeVis} helps to improve the awareness of make-up removal and skincare.

\subsection{Algorithm Robustness Improvement \& Customization}

To reduce the burden on the back-end server, we do not use the mode of machine learning model training for image processing, but also as a result the visual algorithms will have the problem of weak generalization performance. There are various makeup datasets with whole faces (e.g. ~\cite{yan2022beautyrec, 10.1145/3240508.3240618}) that have been used in vision algorithm development, however, to our knowledge, there are no datasets using \textit{EyeVis}, a camera with a special focal length that produces images, have been collected and widely used. Therefore, increasing the diversity of the dataset can also provide more possibilities to improve the robustness of the algorithm. For example, (1) collect more images of Xiqu performers wearing eye makeup with multimodal attributes \rv{(e.g. correspondence between eye make-up and Xiqu acting roles, the degree of opening and closing of the eyes, etc.)}, and (2) invite experts, such as dermatologists, to perform large-scale and more precise image annotation \rv{(e.g. distinguish black face paint and eyelash, distinguish different levels of residue, etc.)}. 

Regarding algorithm customization, we note that it is necessary to study specific operatic characters and develop individualized visualization methods. Even though pink and black face paint is the more common makeup colors, Xiqu performers often adjust the intensity of their makeup according to the stage lighting or the special requirements of their roles, and some local operas use special face paints (e.g. Yingge~\cite{guo2023application}, Tibetan Opera~\cite{calkowski1991day}) that require makeup with multiple complex colors. Generalizing the visual algorithms we have designed to more niche roles and genres will benefit more Xiqu performers. 

\subsection{System Portability}

With the development of mobile phone chips, the response speed of its CPU can also meet the requirements of using mobile phone cameras as hardware for eye-capture tasks has become more common (e.g. (1) eye-tracking in virtual reality~\cite{10.1145/2857491.2857541}, (2) eyelid detection~\cite{fan2021eyelid}, etc.). However, many systems related to skin imaging have yet to apply ubiquitous computing to mobile phones. Compared to professional skin imaging devices for medical use (e.g. VISIA~\cite{goldsberry2014visia}, which contains a desktop operating system with a huge casing that covers the whole face), mobile phones are available to everyone, and the fill light and lens design of \textit{EyeVis} can be taken out and used anywhere and does not take up much space. 

Despite this, \textit{EyeVis} is still a stop-gap measure. As we utilized mobile cameras as image-capturing tools, even if we use a camera lens magnifier, the photos taken by different models of mobile phone cameras will still be slightly different. The choice between building our own camera module to achieve image uniformity across users, or utilizing the computing resources of the phone to mitigate the development costs is a trade-off. \rv{In addition, since each user's facial features are different, it would be possible to personalize our system even more if the eye shields could all be cut to fit different people's faces.} How to reduce the development cost as much as possible while achieving the unity of the system use effect, will be worth studying in future work.

\subsection{Generalizability}
\subsubsection{Wider Area: From Eye to Skin}
Since the launch of the formative study, we focused on the eye area, but in reality, our system can be applied to any skin area other than the eye (e.g. nose, lips). As long as the area can be completely covered by the eye shield, we can still provide stable illumination and high-dimension image quality for image processing. Even if complete coverage by the eye shield is hard, we can fine-tune the focus of the macro camera magnifier and the diameter size of the eye shield to achieve the desired effect.

\subsubsection{Wider Audiences \& Scenarios: From Xiqu Performers to People Who Makeup Frequently}
Our system is not limited to detecting face paints that are used by Xiqu performers. For everyday makeup \& skincare products, some people who makeup frequently also face problems with incomplete removal or wearing. For example, sunscreen, people may not be sure if it has been fully applied to the designated area. While sunscreen has effects to reflect ultraviolet light~\cite{morabito2011review, ngoc2019recent}, the idea of uv simulation can also be applied to detect the level of sunscreen coverage. Therefore, our system can also be applied to everyday makeup \& skincare products and serve people who makeup frequently.

\subsubsection{Wider Usage: Skin Data Collection Tool}
In addition to visualizing and comparing makeup residue, our system can also be used as a skin data collection tool. Skin image databases are used for various scenarios, including clinical diagnosis~\cite{li2021construction, shrivastava2016computer}, skin condition prediction~\cite{10.1145/882262.882344}, racial disparities analysis~\cite{10.1145/3593013.3594114}, etc. Compared to existing skin data collection tools, our system has these advantages: (1) Portable and easy to use, we do not need another customized camera or hardware, and one mobile camera is enough for achieving effects under macro cameras, (2) Cost-effective in centralizing data, unlike devices that collect images locally, our system is directly connected to the remote server, so the remote central server has direct access to the images collected by the systems in each location. Therefore, our system can provide skin imaging developers with the convenience of collecting data and provide some reference values for the advancement of related technologies.
\section{Conclusion and Future Work}

This work emphasized the skin health issues Xiqu performers experience as a result of applying heavy metal-containing face paints for an extended period of time. To investigate the practices and challenges that Xiqu performers face during the skincare process, we conducted an online survey and semi-structured interviews. The findings showed that human-induced skin issues for Xiqu performers were primarily caused by improper makeup removal, particularly from the eyes. To this end, we proposed \textit{EyeVis}, a prototype that can visualize the makeup residue around the eye area and keep tracking the makeup-wearing time of Xiqu performers. The seven-day deployment study verified the usability and long-term results of \textit{EyeVis}. The findings showed that \textit{EyeVis} increases Xiqu performers' awareness of skincare and makeup removal while also boosting their self-assurance and security in skincare. 

In addition to highlighting the significance of proper makeup removal and skincare in the performing arts industry, we also offered insights into the skin health issues experienced by Xiqu performers. For future work, a larger deployment study to validate the effectiveness of \textit{EyeVis} in a larger sample of Xiqu performers will be conducted, and additional features such as personalized skincare recommendations based on skin type and makeup usage will be incorporated into the system. In conclusion, the work offers a promising direction for future investigation into skincare and makeup removal in the performing arts sector and helps to promote and preserve the intangible cultural heritage of practitioners.

\begin{acks}
We thank our reviewers for their constructive feedback and Xiqu performers from Guangdong Cantonese Opera Theatre, Guangzhou Youth Peking Opera Association, Guangdong Dance and Drama College, National Academy of Chinese Theatre Arts, Northern Kunqu Opera Theatre, Suzhou Kunqu Opera Theatre, Changzhou Wuxi Opera Theatre, Sichuan Opera Theatre, and Wenzhou Yue Opera Theatre, for their participation. This work is partially supported by 1) 2024 Guangzhou Science and Technology Program City-University Joint Funding Project (PI: Mingming Fan); 2) 2023 Guangzhou Science and Technology Program City-University Joint Funding Project (Project No. 2023A03J0001) ; 3) Guangdong Provincial Key Lab of Integrated Communication, Sensing and Computation for Ubiquitous Internet of Things (No.2023B1212010007).
\end{acks}

\begin{CJK*}{UTF8}{gbsn}

\bibliographystyle{ACM-Reference-Format}
\bibliography{sample-base}

\end{CJK*}

\appendix
\newpage
\section{Demographic Data of Semi-Structured Interview Participants}
\label{append: demographic}

\begin{table}[ht]
\resizebox{\linewidth}{!}{
\begin{tabular}{@{}cccccccc@{}}
\toprule
\multicolumn{1}{l}{Id} & \multicolumn{1}{l}{Gender} & \multicolumn{1}{l}{Age}  & \multicolumn{1}{l}{Years of Xiqu Makeup Experience} & \multicolumn{1}{l}{Opera Type} & \begin{tabular}[c]{@{}c@{}}Avg  Face Paint Wearing Duration \\ (hour)\end{tabular} & \multicolumn{1}{l}{Avg Face Paint Wearing Frequency} \\ \midrule
1                      & M                          & 15-18                           & 6-10                                                & Cantonese Opera                & 1-2                                                                                & 2-3 times / week                                     \\
2                      & F                          & 26-35                                      & 1-5                                                 & Beijing Opera                  & 1-2                                                                                & 1-2 times / month                                    \\
3                      & F                          & 15-18                            & \textgreater{}10                                    & Beijing Opera                  & 4-6                                                                                & 1-2 times / month                                    \\
4                      & M                          & 18-25                                       & 1-5                                                 & Beijing Opera                  & 2-4                                                                                & 1-2 times / month                                    \\
5                      & F                          & 18-25                                       & 6-10                                                & Yue Opera                      & \textgreater{}6                                                                    & \textgreater 3 times / week                          \\
6                      & F                          & 18-25                                      & 1-5                                                 & Kunqu                          & 2-4                                                                                & once a week                                          \\
7                      & F                          & 18-25                                           & 1-5                                                 & Wuxi Opera                     & 4-6                                                                                & once a week                                          \\
8                      & F                          & \textgreater{}55                           & \textgreater{}10                                    & Beijing Opera                  & 1-2                                                                                & 2-3 times / week                                     \\
9                      & F                          & 36-45                                     & \textgreater{}10                                    & Cantonese Opera                & 4-6                                                                                & once a week                                          \\
10                     & F                          & 15-18                           & 6-10                                                & Cantonese Opera                & 4-6                                                                                & 1-2 times / month                                    \\
11                     & M                          & 36-45                                         & \textgreater{}10                                    & Kunqu                          & \textgreater{}6                                                                    & \textgreater 3 times / week                          \\
12                     & M                          & 26-35                                     & 6-10                                                & Henan Opera                    & 2-4                                                                                & 1-2 times / week                                     \\
13                     & F                          & >55                                   & >10                                                & Cantonese Opera                    & 4-6                                                                                & 2-3 times / week   
\\
14                     & F                          & >55                                    & 6-10                                                & Cantonese Opera                    & 2-4                                                                                & 1-2 times / week       
\\
15                     & M                          & >55                                      & 1-5                                                & Beijing Opera                    & 2-4                                                                                & once a week       
\\

\bottomrule
\end{tabular}
}
\caption{Basic demographic information of the 15 participants in semi-structured interviews}
\label{table: meta}
\Description{Table 3 shows the demographic data of 15 participants who participated in the formative study, the data contains participant ID, gender, age, years of xiqu makeup experience, opera type, average face paint wearing duration, and average face paint wearing frequency.}
\end{table}

\section{Quantitative Comparison of \textit{EyeVis} and \textit{Daily Makeup Removal}}
\label{append: quant}

\begin{table}[tbh!]
\begin{tabular}{|c|c|c|c|c|}
\hline
    & $r_{p\_baseline}$ & $r_{p\_EyeVis\_mean}$ & $r_{b\_baseline}$ & $r_{b\_EyeVis\_mean}$ \\ \hline
P1  & 22.1\%             & 2.8\%                  & 18.4\%             & 3.2\%                  \\ \hline
P2  & 29.7\%             & 5.4\%                  & 26.4\%             & 3.7\%                  \\ \hline
P3  & 25.5\%             & 3.7\%                  & 28.1\%             & 4.6\%                  \\ \hline
P4  & 17.9\%             & 4.4\%                  & 21\%               & 2.5\%                  \\ \hline
P5  & 27.2\%             & 6.9\%                  & 29.1\%             & 6.2\%                  \\ \hline
P6  & 11.4\%             & 1.6\%                  & 8.9\%              & 1.3\%                  \\ \hline
P7  & 14.2\%             & 3.9\%                  & 12.2\%             & 4.1\%                  \\ \hline
P8  & 22.1\%             & 5.2\%                  & 24.7\%             & 6.1\%                  \\ \hline
P9  & 26\%               & 5.4\%                  & 21.4\%             & 4.7\%                  \\ \hline
P10 & 16.4\%             & 3.2\%                  & 13.8\%             & 3.1\%                  \\ \hline
P11 & 9.9\%              & 1.1\%                  & 11.6\%             & 1.2\%                  \\ \hline
P12 & 15.7\%             & 4.4\%                  & 18.2\%             & 2.9\%                  \\ \hline
\end{tabular}
\caption{Ratio data of specified regions of the final taken eye image of each participant's daily makeup removal process (baseline) and using \textit{EyeVis} (mean value of 5 trials). $r_{p\_baseline}$, $r_{b\_baseline}$: ratio of pink \& black face paint area to the whole baseline image. $r_{p\_EyeVis\_mean}$, $r_{b\_EyeVis\_mean}$: ratio of the mean value of pink \& black face paint area to the whole image (using \textit{EyeVis}).}
\Description{Table 4 shows the ratio data of specified regions of the final taken eye image of each participant's daily makeup removal process (baseline) and using EyeVis.}
\end{table}

\end{document}